\documentclass[
twocolumn, preprintnumbers, nofootinbib,amsmath,amssymb]{revtex4}

\usepackage{graphicx}
\usepackage{dcolumn}
\usepackage{bm}
\usepackage{soul}
\usepackage{xcolor}

\newcommand\ee{\end{equation}}
\newcommand\be{\begin{equation}}
\newcommand\eea{\end{eqnarray}}
\newcommand\bea{\begin{eqnarray}}

\def\beq{\begin{equation}}
\def\eeq{\end{equation}}

\newcommand{\bsb}{\boldsymbol}

\newcommand{\q}{{\bsb q}}
\newcommand{\kb}{{\bsb k}}
\newcommand{\x}{{\bsb x}}
\newcommand{\expect}[1]{\left\langle #1 \right\rangle}
\newcommand{\lb}{\ell_{\rm BAO}}
\newcommand{\comment}[1]{}

\renewcommand\[{\left[}

\usepackage[hidelinks]{hyperref} 
\comment{
\hypersetup{
colorlinks=true,
citecolor=DarkBlue,
linkcolor=DarkBlue,
urlcolor=DarkBlue,
}}

\begin{document}

\title{Baryon Acoustic Peak and the Squeezed Limit Bispectrum}

\author{Mehrdad Mirbabayi}
\author{Marko Simonovi\'c}
\author{Matias Zaldarriaga}
\affiliation{Institute for Advanced Study, Einstein Drive, Princeton, NJ 08540, USA}

\begin{abstract}
In the non-relativistic regime, pertinent to the large scale structure of the Universe, the leading effect of a long-wavelength perturbation $\delta(\lambda_L)$ on short distance physics is a uniform acceleration $\propto \lambda_L \delta(\lambda_L)$. Typically, this has no effect on statistical averages at equal time since a uniform acceleration results in a uniform translation --- a reasoning that has been formalized as a ``consistency condition'' on the cosmological correlation functions. This naive expectation fails in the presence of the baryon acoustic feature provided $\lambda_L < \ell_{\rm BAO}$. We derive the squeezed limit of correlation functions in this regime.

\end{abstract}
\maketitle

\noindent

Local experiments performed in a small laboratory cannot reveal the existence of a long-wavelength matter density perturbation $\delta_L(\x,t) = \delta_\q(t) \cos \q\cdot \x$. By equivalence principle, the laboratory and all of its belongings fall with a uniform acceleration $-\nabla \Phi_L(\x_{\rm lab},t)$, where $\Phi_L(\x,t)=-4\pi G a^2 \bar\rho(t)\delta_L(\x,t)/q^2$, and $\bar \rho(t)$ is the mean matter density of the Universe. However, two distant laboratories with separation larger than $1/q$ experience different accelerations. A distant observer sees a clear correlation between the relative motion of the two and the underlying density perturbation.

The small laboratories of the cosmologist, like stars and galaxies, are observed at a single point in their lifespan. Hence, the relative motion of any given pair is impossible to determine. What is possible is to see how the distribution of pairs is correlated with $\delta_L$. In the field of a long-wavelength mode, any galaxy has moved by 
\be
\Delta \x=\delta_q(t) \sin(\q\cdot \x) \, \q/q^2.
\ee
This gives
\be
\label{xi}
\begin{split}
\expect{\delta_g(\frac{\x}{2},t)\delta_g(-\frac{\x}{2},t)}_{\delta_L}&\simeq \xi_g(\x,t)\\[8pt]
+2\delta_q(t) &\sin\left(\frac{\q\cdot\x}{2}\right) \frac{\q}{q^2}\cdot \nabla \xi_g(\x,t),
\end{split}
\ee
where $\xi_g(\x,t)$ is an average 2-point correlation function. Not surprisingly, the distribution of pairs with separation much less than the long wavelength, $\q\cdot\x\ll 1$, is hardly effected by the long mode. The second line would in this case correspond to the effect of living in an over (under) dense Universe. An effect of order $\delta_L x |\nabla \xi_g|$, which for an approximately scale invariant spectrum, $|\nabla \xi_g(\x,t)| \sim \xi_g(\x,t)/x$, is comparable to contributions which are neglected anyway on the right-hand side. 

However, even if $\q\cdot\x\gg 1$, when we do expect the long-wavelength mode to induce a large relative motion, the second line of \eqref{xi} is often negligibly small. Scale invariance now implies that it is of order $\delta_L \xi_g /q x$. A large relative motion is noticeable only if the distribution of pairs has a preferred length scale. One such scale does exist at the baryon acoustic peak. For $x\sim \lb$, 
\be
|\nabla \xi_g(\x,t)|\sim\frac{1}{\sigma}\xi_g(\x,t) \gg \frac{1}{\lb}\xi_g(\x,t),\\[10pt]
\ee
where $\sigma$ is the width of the peak. Correlating \eqref{xi} with $\delta(\q,t)$, we obtain
\be
\begin{split}
\expect{\delta(\q,t) \delta_g(\frac{\x}{2},t)\delta_g(-\frac{\x}{2},t)}&\\[8pt]
\simeq 2 P_{\rm lin}(q,t) \sin\left(\frac{\q\cdot\x}{2}\right)& \frac{\q}{q^2}\cdot \nabla \xi_g(\x,t),
\end{split}
\ee
where $P_{\rm lin}(q,t)$ is the linear matter power spectrum and the corrections are sub-dominant as long as $q \ll \sigma^{-1}$. In momentum space, we have for $k\gg q$
\be\label{bi}
\begin{split}
&\expect{\delta(\q,t) \delta_g(\kb-\q/2,t)\delta_g(-\kb-\q/2,t)}\simeq P_{\rm lin}(q,t) \\[8pt]
&\frac{\q\cdot \kb}{q^2} \left[P_g(|\kb-\q/2|,t)-P_g(|\kb+\q/2|,t)\right],
\end{split}
\ee
\comment{where $P_g^w(k,t)$ is the power in the acoustic peak: the Fourier transform of $\xi_g$ after the subtraction of a scaling background. In terms of the full power, the right-hand side can be written as
\be
\frac{\q\cdot \kb}{q^2} P_{\rm lin}(q,t)\left[P_g(|\kb-\q/2|,t)-P_g(|\kb+\q/2|,t)\right],
\ee}
which can be generalized to higher point statistics. The above difference of power spectra is normally assumed to be of order $q$ in the derivation of squeezed limit consistency conditions. This has led to the conclusion that the $1/q$ contribution to the squeezed limit bispectrum vanishes at equal times (see e.g. \cite{Kehagias:2013yd, Peloso:2013zw, Creminelli:2013mca}). Evidently, this is not the case in the presence of the acoustic feature since $P_g(k,t)$ has an oscillatory component with period $2\pi \lb^{-1}$.

Our derivation has two underlying assumptions: the equivalence principle, and the absence of primordial local non-Gaussianity. Therefore, it holds beyond the standard perturbation theory, and to all orders in the displacement field $\Delta \x$ (see e.g. \cite{Creminelli:2013poa}). Intuitively, the above result describes how galaxy pairs, which are more likely to be found at distance $\lb$, are moved to larger or smaller separations in the presence of a long-wavelength mode. When averaged over the long modes, the result would be a slightly wider acoustic peak. This broadening can be obtained, by using the squeezed limit formula \eqref{bi} in the one-loop calculation of the correlation function, or just expanding \eqref{xi} to higher orders in displacements and averaging \cite{Sugiyama:2013gza, Senatore:2014via}. As another practical application, equation \eqref{bi} can potentially be used to determine bias coefficients for matter tracers by replacing $\delta(\q,t)$ in terms of the tracer density contrast $\delta_g(\q,t)$.

{\em Acknowledgements.---} We thank T. Baldauf and P. Creminelli for useful discussions. M.M. is supported by NSF Grants PHY-1314311 and PHY-0855425. M.S. acknowledges support from the Institute for Advanced Study. M.Z. is supported in part by the NSF grants AST- 0907969, PHY-1213563 and AST-1409709.

\end{document}